\documentclass[PAPER, atlasdraft=false, coverpage=false, USenglish, cernpreprint, texmf, orcidlogo]{atlasdoc}
\usepackage{atlaspackage}
\usepackage{atlasbiblatex}

\usepackage{atlasphysics}
\graphicspath{{logos/}{figures/}}

\usepackage{ANA-TOPQ-2025-15-PAPER-defs}

\AtlasTitle{Observation of single top-quark production in the $s$-channel in proton--proton collisions at $\sqrt{s}=13$ TeV with the ATLAS detector}

\AtlasAbstract{%
The first single-experiment observation of single top-quark production in the $s$-channel is presented. Measurements of this process are essential to fully characterize the electroweak interactions of the top quark and probe for new physics, however its highly suppressed cross-section has so far caused it to evade direct detection at the Large Hadron Collider. The analysis uses proton--proton collision data collected at a center-of-mass energy of $\sqrt{s}=13$~TeV with the ATLAS detector, corresponding to an integrated luminosity of $140$~fb$^{-1}$. The selected events contain exactly one electron or muon, missing transverse momentum, and exactly two $b$-quark-initiated jets. A Graph Neural Network is utilized to separate the signal from the top-quark pair, $W$+jets, and $t$-channel single top-quark backgrounds. A likelihood fit to the network output yields an observed (expected) signal significance of $5.1$~($5.2$)~standard deviations. The measured cross-section of \mbox{$\sigma_{t\bar{b}+\bar{t}b} = 10.8 \pm 1.5 \text{ (stat.)} \pm 1.6 \text{ (syst.)}\,\si{\pico\barn}$} is consistent with the Standard Model prediction.
}

\author{The ATLAS Collaboration}

\AtlasRefCode{TOPQ-2025-15}

\PreprintIdNumber{CERN-EP-2026-265}

\AtlasJournal{Phys.\ Rev.\ Lett.}
\AtlasCoverEgroupAnalysisTeam{atlas-topq-2025-15-analysis-team@cern.ch}

\hypersetup{pdftitle={ATLAS document},pdfauthor={The ATLAS Collaboration}}

\begin{document}

\maketitle

%
%

%

%
%
%

Measurements of single top-quark production are essential to fully characterize top quark production via electroweak interactions. In the Standard Model, single top-quark production proceeds via three mechanisms: the \tchannel and \schannel production, distinguished by the virtuality of the $W$ boson, and associated $tW$ production.
Single-top-quark processes are directly sensitive to the Cabibbo-Kobayashi-Maskawa (CKM) matrix element $V_{tb}$ and sensitive
to new resonances such as $W'$ or charged Higgs bosons present in various physics scenarios beyond the Standard Model~\cite{Tait:2000sh,Boos:2006xe} or in effective field theories~\cite{Zhang:2016omx}.

\begin{figure}[!ht]
\centering
\includegraphics{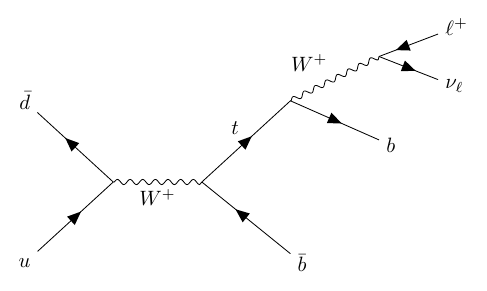}
\caption{Representative Feynman diagram of single top-quark production via the \schannel exchange of a virtual \wboson at leading-order in perturbation theory, followed by the subsequent semileptonic top-quark decay.}
\label{fig:feynman_schannel_decay}
\end{figure}

While the \tchannel and $tW$ modes are firmly established~\cite{TOPQ-2023-19, CMS-TOP-17-023, CMS-TOP-21-010} at the Large Hadron Collider (LHC)~\cite{Evans:2008zzb}, the \schannel remains the last direct top-quark production mechanism to evade observation at the LHC\@. At the Tevatron, however, the \schannel was observed by combining data from the CDF and D\O\ collaborations~\cite{CDF:2014uma} in proton--anti-proton collisions at a center-of-mass energy of \SI{1.96}{\TeV}. At the LHC, the \schannel cross-section is heavily suppressed because the leading-order (LO) process, illustrated in Figure~\ref{fig:feynman_schannel_decay}, requires a sea antiquark in the initial state, although higher-order corrections involving quark-gluon initial states provide substantial enhancements~\cite{Heim:2009ku, Liu:2018gxa}. The ATLAS Collaboration previously reported evidence for this process at center-of-mass energies of \SI{8}{\TeV}~\cite{TOPQ-2015-01} and \SI{13}{\TeV}~\cite{TOPQ-2018-04}, achieving observed~(expected) significances of $3.2$~($3.9$) and $3.3$~($3.9$) standard deviations, respectively. Additionally, searches by the CMS Collaboration at \SI{7}{\TeV} and \SI{8}{\TeV}~\cite{CMS-TOP-13-009} yielded an observed significance of $2.5$ standard deviations.

The predicted cross-section of the signal \schannel (\tbtb) production at a center-of-mass energy of \SI{13}{\TeV} is $\sigma(t\bar{b}+\bar{t}b) = 11.07 ^{+0.19}_{-0.18}\,\si{\pico\barn}$ as computed at next-to-next-to-leading-order (NNLO) in Quantum Chromodynamics (QCD)~\cite{Liu:2018gxa} using a top-quark mass of \SI{172.5}{\GeV} and the PDF4LHC21 parton distribution function (PDF) set~\cite{Ball:2022oua}. The uncertainty includes contributions from the choice of PDFs, and the QCD renormalization scale \muR and factorization scale \muF.

This Letter reports the first observation of single top-quark \schannel production at the LHC\@. The measurement is performed using the full Run~2 data of proton--proton collisions at a center-of-mass energy of \SI{13}{\TeV} collected by the ATLAS detector, corresponding to an integrated luminosity of \SI{140}{\ifb}. While utilizing the same data sample as the previous ATLAS analysis~\cite{TOPQ-2018-04}, this measurement achieves substantially higher sensitivity. The improvement is driven by a broadened kinematic acceptance that significantly increases the number of available signal events, improved $b$-tagging algorithms, refined theoretical modeling, and the introduction of a Graph Neural Network (GNN) to extract the signal from the dominant top-quark pair (\ttbar), \wjets, and \tchannel single top-quark backgrounds, and other minor background contributions.

The ATLAS experiment~\cite{PERF-2007-01} at the LHC is a multipurpose particle detector with a forward--backward symmetric cylindrical geometry and a near \(4\pi\) coverage in solid angle.\footnote{\AtlasCoordFootnote} It consists of an inner tracking detector surrounded by a thin superconducting solenoid, electromagnetic and hadronic calorimeters, and a muon spectrometer.
A two-level trigger system was used to select events~\cite{TRIG-2019-04}.
Events were selected using a combination of single-electron and single-muon triggers with varying transverse momentum thresholds and isolation requirements~\cite{TRIG-2018-05, TRIG-2018-01}.
A software suite~\cite{SOFT-2022-02,TopCPToolkit} is used in data simulation, in the reconstruction and analysis of real and simulated data, in detector operations, and in the trigger and data acquisition systems of the experiment.

Sets of simulated events from signal and background processes were produced with Monte Carlo (MC) based event generator programs to model the physics processes. After event generation, the response of the ATLAS detector was simulated using the \GEANT toolkit~\cite{Agostinelli:2002hh} with a full detector model~\cite{SOFT-2010-01} or a fast simulation~\cite{SIMU-2018-04} that employed a parameterization of the calorimeter response. To account for additional proton--proton interactions in the same and neighboring bunch crossings (pileup), inelastic events were simulated using a mix of \EPOS[2.0.1.4]~\cite{Werner:2005jf} with the \EPOS LHC tune~\cite{Pierog:2013ria} and \PYTHIA[8.308]~\cite{Bierlich:2022pfr} with the A3 tune~\cite{ATL-PHYS-PUB-2016-017} and the \NNPDF[2.3lo] PDF set~\cite{Ball:2012cx}, and overlaid~\cite{SIMU-2020-01} on the hard-scattering events.
Samples of simulated events involving top quarks were simulated using the \POWHEGBOX[v2]~\cite{Nason:2004rx,Frixione:2007nw,Frixione:2007vw,Alioli:2009je,Alioli:2010xd,Re:2010bp,Frederix:2012dh} matrix element generator with the top-quark mass set to \SI{172.5}{\GeV}. All processes were calculated at next-to-leading-order (NLO) in QCD and used the \NNPDF[3.0nlo] PDF set~\cite{Ball:2014uwa} with the five-flavor scheme, except for single top-quark production in the \tchannel, which was simulated with the \NNPDF[3.0nlo\_nf$_4$] PDF set in the four-flavor scheme.
Parton showers, hadronization, and the underlying event were modeled using \PYTHIA[8.230]~\cite{Sjostrand:2007gs} with the A14~\cite{ATL-PHYS-PUB-2014-021} set of tuned parameters and the \NNPDF[2.3lo] PDF set~\cite{Ball:2012cx}. When generating $tW$ events, the diagram-removal scheme~\cite{Frixione:2008yi} was employed to treat the interference with \ttbar\ production~\cite{ATL-PHYS-PUB-2016-020}. To improve the modeling of the top-quark transverse momentum distribution~\cite{TOPQ-2024-12}, the simulated \ttbar events were corrected via a reweighting procedure to match differential cross-section predictions at NNLO in QCD and NLO in electroweak (EW) accuracy~\cite{Czakon:2017wor}. The decays of bottom and charm hadrons were simulated using the \EVTGEN~1.6.0 program~\cite{Lange:2001uf}.
The production of $W$ and $Z$ bosons in association with jets was simulated with the \SHERPA[2.2.11] generator~\cite{Bothmann:2019yzt} and for diboson production ($WW$, $WZ$, and $ZZ$) with the \SHERPA[2.2.14] generator. In this setup, NLO-accurate matrix elements for up to two (one) partons and LO-accurate matrix elements for up to five (three) partons for $W/Z$+jets (diboson) production are calculated with the \textsc{Comix}~\cite{Gleisberg:2008fv} and \OPENLOOPS~\cite{Buccioni:2019sur,Cascioli:2011va,Denner:2016kdg} libraries. The default \SHERPA parton shower~\cite{Schumann:2007mg} and the cluster hadronization model~\cite{Winter:2003tt} are used. The \MEPSatNLO prescription~\cite{Hoeche:2011fd,Hoeche:2012yf,Catani:2001cc,Hoeche:2009rj} is used to remove overlap between partonic configurations generated by the matrix element and by parton shower evolution. The generation employed the dedicated set of tuned parameters developed by the \SHERPA authors and the \NNPDF[3.0nlo] PDF set. The $W$+jets sample is separated into events with heavy-flavor (HF) $c$ or $b$ hadrons and the remaining light-flavor (LF) component. Processes involving the production of a Higgs boson are neglected due to their much smaller production cross-sections. Multijet events with fake electrons or non-prompt electrons from hadron decays are described with simulated dijet samples, while non-prompt muons are estimated from collision data by applying the nominal selection with inverted or modified muon isolation requirements. The dijet event sample was generated at LO in QCD using \PYTHIA[8.186]~\cite{Sjostrand:2007gs} with the \NNPDF[2.3lo] PDF set and the A14 set of tuned parameters.

The partonic final state of the \schannel signal process, illustrated in Figure~\ref{fig:feynman_schannel_decay}, comprises a charged lepton, a neutrino, and two $b$-quarks, and is selected by identifying corresponding objects measured in the detector.
The presence of a high-\pT neutrino is indicated by large missing transverse momentum.

Events must contain a primary vertex, defined by the highest sum of squared transverse momenta of at least two associated tracks~\cite{IDTR-2021-01}.
Electron candidates are reconstructed from Inner Detector (ID) tracks matched to energy clusters in the electromagnetic calorimeter. A likelihood-based discriminant is constructed and the \textit{Tight} identification criteria is applied to reject non-prompt electrons~\cite{EGAM-2021-01}.
Muon candidates are reconstructed by combining ID and muon spectrometer tracks and identified using the \textit{Medium} quality requirements that reject fake muons~\cite{PERF-2015-10, MUON-2018-03}.
Leptons must be consistent with this vertex via impact-parameter requirements, and are further subject to isolation criteria optimized to reject non-prompt and fake candidates.
Scale factors correct simulation efficiencies to match measured efficiencies for electron and muon triggers, and reconstruction, identification, and isolation criteria.
The leptons must satisfy transverse momentum $\pt>\SI{10}{\GeV}$ and $|\eta|<2.5$.
Jets are reconstructed from particle-flow objects~\cite{PERF-2015-09} with the anti-$k_t$ algorithm~\cite{Cacciari:2008gp,Fastjet} using a radius parameter $R=0.4$.
The energy scale of jets is calibrated by applying several simulation-based corrections and techniques
correcting for differences between simulation and data~\cite{JETM-2018-05}.
The jets must satisfy $\pt>\SI{30}{\GeV}$ and $|\eta|<2.5$.
Pileup suppression is performed via the neural-network-based jet vertex tagger, similar to Ref.~\cite{PERF-2014-03}.
Jets containing \(b\)-hadrons are identified (\(b\)-tagged) with the \texttt{GN2} algorithm,
which uses a graph transformer architecture~\cite{FTAG-2023-05}.
A working point (WP) with \SI{85}{\%} $b$-tagging
efficiency in dileptonic \ttbar events is used. Differences between data and simulation are corrected with scale factors derived from \ttbar and $Z$+jets events, and additional MC-to-MC scale factors are applied to account for differences between the parton-shower and hadronization models across the various event generators.
An overlap-removal procedure is applied to prevent double-counting of objects satisfying multiple reconstruction criteria~\cite{TOPQ-2019-02}.
The missing transverse momentum $\vec{p}_\mathrm{T}^\mathrm{miss}$ is reconstructed as the negative vector sum of the \pT of the reconstructed leptons and jets, and tracks associated with the primary vertex but not matched to any reconstructed object~\cite{JETM-2020-03}. Its magnitude is denoted by \met.

Selected events are required to contain exactly one charged lepton $\ell$ (electron or muon, including those from leptonic $\tau$ decays) with $\pt>\SI{28}{\GeV}$ and are categorized into a signal region (SR) and several orthogonal control regions (CRs) using additional selection criteria. Events in the SR must have exactly two jets, both of which need to be $b$-tagged jets, to suppress semileptonic \ttbar events. To suppress multijet background and poorly reconstructed leptons, events are required to have $\met > \SI{30}{\GeV}$ and a $W$~boson transverse mass $\mtw > \SI{30}{\GeV}$.\footnote{\mtwFootnote} In addition, a requirement is made in the two-dimensional $\pT(\ell)$--$|\Delta\phi(j_1,\ell)|$ plane by selecting a triangular region defined by $\pT(\ell) > \SI{40}{\GeV} \cdot |\Delta\phi(j_1,\ell)|/\pi$~\cite{TOPQ-2019-02}, where $j_1$ is the jet with the largest \pT.
The \ttbar CR is defined by requiring one additional non-$b$-tagged jet. The $W$+jets CR is defined by requiring exactly two jets, both of which must satisfy a looser \SI{90}{\%} $b$-tagging efficiency WP, while at least one of them fails the nominal \SI{85}{\%} WP\@. This specific $b$-tagging requirement creates a phase space orthogonal to the SR that maintains a heavy-flavor composition representative of the \wjets background in the SR, while being depleted of true $b$-jets from \ttbar and the signal process. Three multijet CRs are defined: two electron multijet CRs, separated into barrel ($|\eta| < 1.37$) and endcap ($|\eta| > 1.52$) electrons, obtained by inverting the SR \met requirement, and one muon multijet CR, obtained by inverting the SR triangular requirement. A summary of the expected event yields for the non-multijet regions is given in \Tab{\ref{tab:yields_SR_CR}}.

\begin{table}[htbp]
\caption{Pre-fit expected and observed number of events in the SR, the \ttbar CR, and the $W$+jets CR\@. The uncertainties in the expected yields include all systematic and MC statistical uncertainties.}
\label{tab:yields_SR_CR}
\centering
\begin{tabular}{l|S|S|S}
\toprule
& {SR} & {\ttbar CR} & {$W$+jets CR} \\
\midrule
$t\bar{b}+\bar{t}b$   & 12900 \pm 700 & 6400 \pm 600 & 2010 \pm 140 \\
$tq+\bar{t}q$         & 95000 \pm 8000 & 72000 \pm 7000 & 24500 \pm 1700 \\
$tW+\bar{t}W$         & 25400 \pm 2900 & 54000 \pm 5000 & 16400 \pm 1600 \\
$t\bar{t}$            & 560000 \pm 60000 & 1350000 \pm 100000 & 146000 \pm 15000 \\
$W$+HF                & 101000 \pm 11000 & 77000 \pm 15000 & 133000 \pm 22000 \\
$W$+LF                & 6400 \pm 2300 & 3900 \pm 2200 & 35000 \pm 11000 \\
$VV, Z$+jets          & 21900 \pm 4100 & 16800 \pm 3600 & 27100 \pm 6000 \\
Multijet             & 30600 \pm 1600 & 21200 \pm 2800 & 20400 \pm 1300 \\
\midrule
Total                 & 850000 \pm 60000 & 1600000 \pm 100000 & 404000 \pm 33000 \\
\midrule
Data                  & {858387} & {1580729} & {404781} \\
\bottomrule
\end{tabular}
\end{table}

The multijet background templates are derived using a combination of simulated dijet events and data-driven techniques~\cite{TOPQ-2015-05,TOPQ-2019-02,TOPQ-2023-44}. The background from jets misidentified as electrons is modeled using simulated dijet events, wherein jets depositing a large fraction of their energy in the electromagnetic calorimeter are treated as electron candidates. The background from non-prompt muons is estimated directly from collision data using a multijet-enriched sample. These methods provide the shape of the multijet distributions but do not determine the overall normalization. A preliminary multijet fit in the corresponding CRs is used to estimate the normalization for the pre-fit event yield.

To isolate the small \schannel signal from overwhelming backgrounds from \ttbar and \wjets production, a Graph Neural Network~\cite{battaglia2018} is employed. The GNN models each collision event as a fully connected graph, defining the lepton, the two $b$-tagged jets, and the \met as individual nodes. Each node is parameterized by its transverse momentum \pt, pseudorapidity $\eta$, and azimuthal angle $\phi$, alongside object-specific auxiliary features: the charge for leptons and the \texttt{GN2} binned $b$-tagging score~\cite{FTAG-2023-05}, used to categorize jets into exclusive intervals corresponding to successive $b$-tagging efficiency working points, providing additional flavor-discrimination information beyond a single fixed operating point.
Edges are formed between all possible pairs of nodes to encode their kinematic relationships, utilizing $\Delta\phi$, $\Delta\eta$, $\Delta R$, $\Delta\pt$, and the invariant mass $m_\text{inv}$ of the pair. To compute these geometric and mass features for edges connected to the \met node, the missing transverse momentum is treated as a massless four-vector with $\eta = 0$. Finally, the scalar sum of the transverse momenta of all objects \Ht is included as a global graph attribute. By operating directly on these geometric and kinematic properties, the GNN efficiently encodes the complete event topology, yielding significantly enhanced signal-to-background discrimination compared with previous methods~\cite{TOPQ-2018-04,TOPQ-2014-03}. The network architecture comprises four consecutive message-passing blocks. Within each block, the edge, node, and global feature representations are sequentially updated using Multi-Layer Perceptrons, each configured with four hidden layers of 64 nodes.

The GNN is implemented in \textit{TensorFlow}~\cite{tensorflow} using the \textit{Graph Nets} library~\cite{battaglia2018}. It is trained as a binary classifier to separate the \schannel signal from all background processes by minimizing a binary cross-entropy loss function. The network hyperparameters, including the learning rate and dropout probability, are optimized empirically to maximize the signal-to-background discrimination, with the model performance found to be robust across a wide range of parameter variations. To prevent the network from becoming biased towards the dominant background classes, the overall signal and combined background events are normalized to a one-to-one ratio during the loss calculation. To ensure statistically independent training and testing data samples, a two-fold cross-validation strategy is employed: two independent models are trained separately on events with even and odd event numbers. During inference, each model is applied to the orthogonal subset to preclude overtraining.

Both the overall normalization and the kinematic shapes of the signal and background expectations are affected by multiple systematic uncertainties. These are generally classified into two distinct categories, experimental uncertainties accounting for detector resolution and calibration limitations and modeling uncertainties covering effects related to the theoretical description of the simulated events.

The combined 2015--2018 data sample carries an integrated luminosity uncertainty of 0.83\%~\cite{DAPR-2021-01}. Experimental uncertainties in electrons and muons include the uncertainties on the scale factors correcting for trigger, reconstruction, identification, and isolation efficiency discrepancies between data and simulation~\cite{MUON-2018-03, EGAM-2021-02, MUON-2022-01}, alongside momentum scale and resolution variations.
Jet energy scale (JES) and resolution (JER) uncertainties~\cite{JETM-2018-05} are decomposed into uncorrelated nuisance parameters (NPs) encompassing detector and pileup modeling, single-particle response, flavor composition, and residual differences that account for differences between reconstruction algorithms. Additional experimental uncertainties account for the jet-vertex-tagger efficiency~\cite{PERF-2014-03}, corrections that match the simulated pileup distribution to data, and potential mismodelling of the \met soft-track component~\cite{PERF-2016-07}.
Uncertainties in the \btag requirement are factorized into components for the tagging efficiency of true \bjets and the mistag rates of \(c\)- and light-flavor jets~\cite{FTAG-2023-10,FTAG-2024-07,FTAG-2019-02}. The multijet background shape uncertainty is derived by varying the jet-electron and non-prompt-muon selection criteria. Statistical uncertainties due to the limited number of simulated MC events are accommodated using independent per-bin NPs following the Barlow-Beeston method~\cite{Barlow:1993dm}, with the signal treated as uncorrelated with the backgrounds.

Theoretical modeling uncertainties are assessed for all simulated processes. PDF uncertainties are evaluated using the PDF4LHC15 set~\cite{Butterworth:2015oua} for all processes, except for the \twchannel background, which uses the PDF4LHC21 set~\cite{Ball:2022oua}. Consequently, the PDF uncertainties are treated as correlated across all processes except for the \twchannel. The impacts of missing higher-order corrections are estimated by independently varying the renormalization and factorization scales \muR and \muF by factors of $0.5$ and $2.0$. Normalizations for background processes not freely floating in the fit are constrained by \textit{a priori} theoretical cross-section uncertainties of \SI{3.5}{\percent} for the \twchannel~\cite{Kidonakis:2021vob,ATL-PHYS-PUB-2025-035}, \SI{2.0}{\percent} for the \tchannel~\cite{Campbell:2020fhf, ATL-PHYS-PUB-2025-035}, \SI{16}{\percent} for \zjets~\cite{TOPQ-2018-07}, and \SI{13}{\percent} for diboson production~\cite{TOPQ-2018-07}. Additionally, a \SI{15}{\percent} uncertainty is assigned to the relative fraction of $W+b$ and $W+c$ events within the heavy-flavor \wjets background. This induces a \SI{10}{\percent} variation in the \wjets CR due to its larger $W+c$ fraction; consequently, larger prior uncertainties on the $W+b$ vs $W+c$ fraction would be strongly constrained by this region.
Parton shower, hadronization and underlying event uncertainties are estimated by comparing the nominal \POWHEGBOX[v2]~+~\PYTHIA[8.230] setup with a \POWHEGBOX[v2]~+~\HERWIG~\cite{Bahr:2008pv,Bellm:2015jjp} configuration, using \HERWIG[7.2.1] for \ttbar and \twchannel, and \HERWIG[7.3.1] for \tchannel and \schannel production. Uncertainties related to NLO matrix-element matching follow \Refn{\cite{Hoche:2021mkv}}, and the \hdamp parameter uncertainty in \ttbar and $tW$ production is probed by varying \hdamp from its nominal value of $1.5 \times m_t$ to $3 \times m_t$.
Systematic variations covering the reweighting of \ttbar events to NNLO QCD and NLO EW accuracy include NNLO scale choices, the modeling of the photon content of the proton evaluated by comparing different quantum-electrodynamic (QED) PDF sets (\NNPDF[3.0 QED]~\cite{Ball:2014uwa} and LUXqed~\cite{Bertone:2017bme,Manohar:2017eqh,Manohar:2016nzj}), reweighting variable order, and differences between fixed-order and parton-shower modeling for the top-quark \pt and $m_{\ttbar}$ distributions.
Initial- and final-state radiation (ISR/FSR) uncertainties are assessed via the A14 tune's \texttt{Var3c} parameter~\cite{ATL-PHYS-PUB-2014-021}. Finally, the interference ambiguity between $tW$ and \ttbar is evaluated by substituting the nominal diagram-removal scheme with a diagram-subtraction approach~\cite{Frixione:2008yi}. This variation accounts for the theoretical ambiguity in defining the background processes at next-to-leading order.

To measure the \schannel production cross-section and determine the signal significance, a binned profile maximum-likelihood fit is performed simultaneously across the SR and the orthogonal CRs. The fit is implemented using the \TRExFitter framework~\cite{TRExFitter}. The likelihood function is constructed as a product of Poisson probability terms over all bins of the fitted distributions. The binning of the fitted observables is optimized to maximize the expected signal significance. The parameter of interest is the signal strength $\mu$, defined as the ratio of the measured \schannel cross-section to its Standard Model prediction. The distribution of the GNN output score serves as the fitted observable in both the SR and the \wjets CR\@. In the \ttbar CR, the scalar sum of the transverse momenta of all kinematic objects \Ht is used as the discriminating variable. The multijet CRs are included as single-bin yields.
To accurately control the major backgrounds directly from data, the normalizations of the \ttbar background, the heavy- and light-flavor \wjets backgrounds, and the three multijet background components (barrel electron, endcap electron, and muon) are left as free parameters in the fit, alongside the \schannel signal strength. Systematic uncertainties are incorporated as NPs constrained by probability density functions multiplied into the likelihood, using standard Gaussians for experimental and theoretical uncertainties, and Poisson terms for Monte Carlo statistical uncertainties. The NPs are normalized such that a unit variation corresponds to the pre-fit magnitude of each respective uncertainty, allowing the fit to systematically adjust the background and signal expectations to best match the data.

\begin{figure}[ht]
\centering
\includegraphics[width=0.5\textwidth]{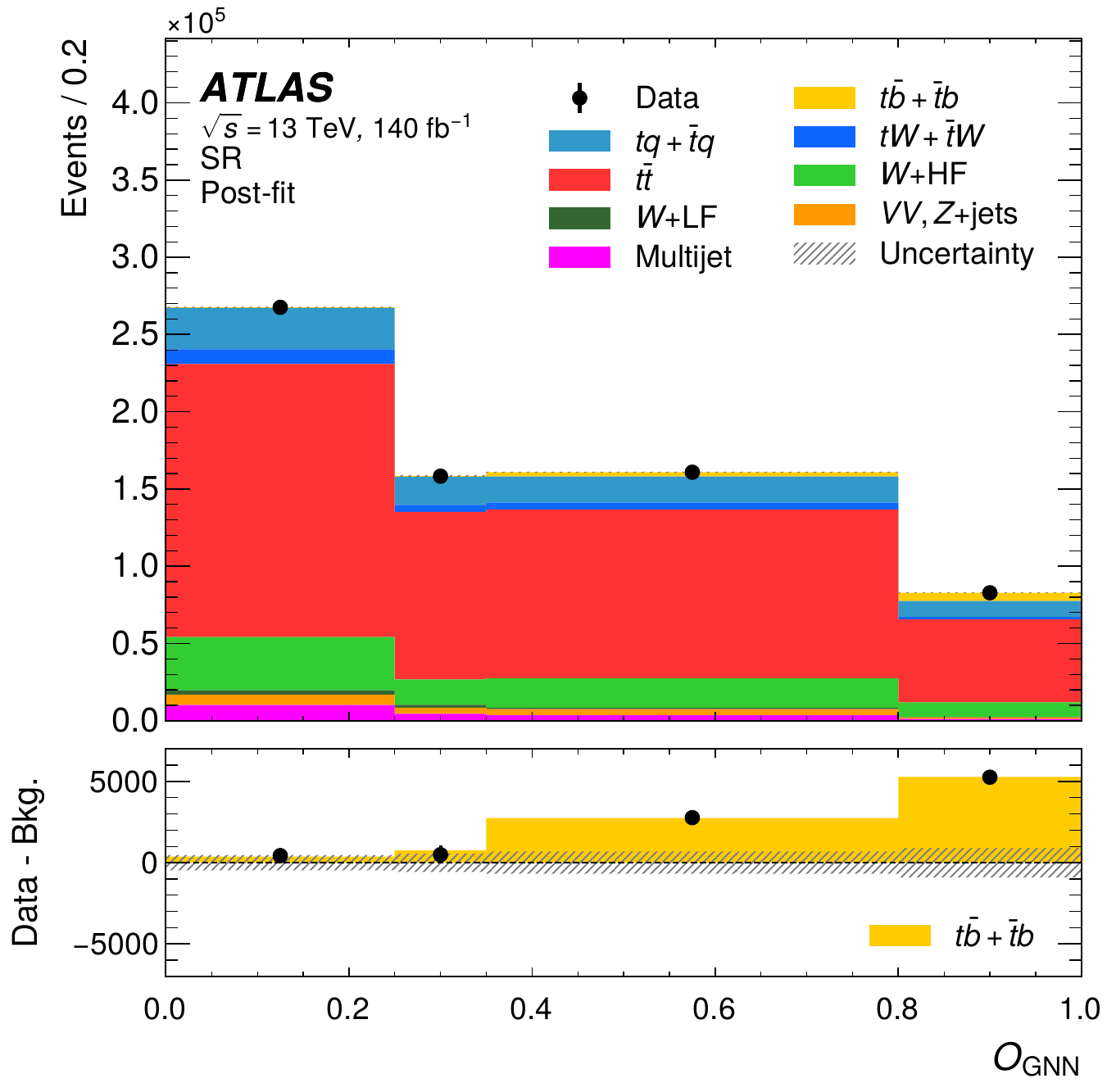}
\caption{Post-fit distribution of the GNN in the signal region. The event yields are scaled to a reference bin width of $0.2$ to reflect the event density across variable-width bins. The shaded band represents the uncertainty in the total prediction after the fit. The small contribution at the top of the stack is the \schannel signal. The lower panel shows the background-subtracted data overlaid with the post-fit signal contribution, with the post-fit uncertainty in the background contribution represented by the shaded area around zero.}
\label{fig:sr_postfit}
\end{figure}

The post-fit GNN output distribution in the SR is shown in Figure~\ref{fig:sr_postfit}, demonstrating agreement between the data and the background-plus-signal prediction, with a goodness-of-fit value of $97\%$ evaluated using the saturated model~\cite{Cousins:2018tiz}. This value is derived from a likelihood-ratio test that compares the nominal fit to a saturated model designed to perfectly describe the observed data, yielding a test statistic that asymptotically follows a $\chi^2$ distribution. This high probability confirms that shifts and constraints on the NPs remain well-behaved and within their expected bounds. The measured post-fit normalization factors are $\mu(\ttbar) = 1.00 \pm 0.07$, $\mu(W\text{+HF}) = 1.1 \pm 0.2$, and $\mu(W\text{+LF}) = 1.2 \pm 0.7$, while the multijet background factors are $0.7 \pm 0.2$ for barrel electrons, $0.96 \pm 0.08$ for endcap electrons, and $0.7 \pm 0.5$ for muons. The significance of the signal is determined by performing a single-sided hypothesis test against the background-only hypothesis ($\mu = 0$). This is evaluated using a test statistic based on the profile likelihood ratio, from which the corresponding significance is extracted using the asymptotic approximation~\cite{Cowan:2010js}. The fit yields an observed~(expected) significance of $5.1$~($5.2$) standard deviations, establishing the observation of \schannel single top-quark production. The inclusive \schannel production cross-section is measured to be \mbox{$\sigma_{t\bar{b}+\bar{t}b} = 10.8 \pm 1.5 \text{ (stat.)} \pm 1.6 \text{ (syst.)}\,\si{\pico\barn}$}, consistent with the Standard Model expectation.

\begin{table}[ht]
\caption{Breakdown of the total fractional uncertainty in the measured \schannel production cross-section $\sigma_{t\bar{b}+\bar{t}b}$, given in \%. Individual nuisance parameter impacts are derived from the fit covariance matrix and summed in quadrature to evaluate the uncertainties for nuisance parameter groups and the total systematic uncertainty~\cite{Pinto:2023yob}. The statistical uncertainty of the data is extracted by subtracting the total systematic variance from the total variance of the profile-likelihood fit.}
\label{tab:systematics_realistic_grouped}
\begin{center}
\begin{tabular}{lc}
\toprule
Uncertainty group     & $\Delta\sigma_{t\bar{b}+\bar{t}b}/\sigma_{t\bar{b}+\bar{t}b}$ [\%] \\
\midrule
Background modeling                     & $11.6$ \\
Signal modeling                         & $4.1$ \\
Parton distribution functions           & $1.5$ \\
Jet calibration                         & $7.9$ \\
$b$-tagging efficiency: $b$-jets        & $2.8$ \\
$b$-tagging efficiency: $c$-jets        & $0.6$ \\
$b$-tagging efficiency: light-jets      & $0.9$ \\
Integrated luminosity                   & $1.0$ \\
Missing transverse momentum             & $0.8$ \\
Lepton calibration                      & $0.3$ \\
MC statistical uncertainty              & $2.7$ \\
\midrule
Total systematic uncertainty            & $15.2$ \\
Statistical uncertainty                 & $13.5$ \\
\midrule
Total uncertainty                       & $20.4$ \\
\bottomrule
\end{tabular}
\end{center}
\end{table}

As detailed in Table~\ref{tab:systematics_realistic_grouped}, systematic uncertainties yield a total fractional uncertainty in the cross-section of approximately 15\%, similar to the statistical uncertainty of the data of about 14\%. The leading systematic contributions arise from background modeling, driven primarily by the theoretical ambiguity in $tW$ and $t\bar{t}$ interference and $W$+jets modeling, and jet-related calibrations. Additional notable impacts originate from signal modeling and $b$-tagging uncertainties, whereas contributions from luminosity, missing transverse momentum, and leptons are comparatively minor. While the baseline uncertainty breakdown is derived from the fit covariance matrix, to compare to the technique used in the previous analysis~\cite{TOPQ-2018-04}, the statistical and systematic uncertainties can be evaluated by fixing all systematic NPs to their post-fit values. This alternative method yields a statistical uncertainty of $\pm 0.7\,\si{\pico\barn}$ and a systematic uncertainty of $\pm 2.1\,\si{\pico\barn}$.

The significantly improved sensitivity of this measurement culminates in the first observation of single top-quark \schannel production at the LHC\@. This milestone is driven by major methodological advancements compared with the previous ATLAS search~\cite{TOPQ-2018-04}, which also analyzed the full Run~2 data. Specifically, a broadened kinematic acceptance increases the number of selected signal events by a factor of approximately~three, while the introduction of a GNN provides superior signal-to-background separation. To measure this improvement, the statistical separation power is evaluated using the approximate significance $S/\sqrt{B}$, where $S$ and $B$ are the expected signal and background yields. The inclusive approximate expected significance ($S/\sqrt{B}$) in the signal region improves from approximately $12$ in the previous search to $14$. Furthermore, in the most sensitive regions capturing about $40\%$ of the signal events, this increases from $13$ to $20$. These techniques, combined with improved $b$-tagging algorithms, improved calibrations, and refined theoretical modeling of both the signal and backgrounds, overcome the immense background challenges that have previously hindered direct detection at these collision energies.

In summary, using ATLAS $pp$ collision data at $\sqrt{s}=13$~TeV, the \schannel process is established with an observed~(expected) significance of $5.1$~($5.2$) standard deviations. The measured inclusive cross-section of \mbox{$\sigma_{t\bar{b}+\bar{t}b} = 10.8 \pm 1.5 \text{ (stat.)} \pm 1.6 \text{ (syst.)}\,\si{\pico\barn}$} is in good agreement with the Standard Model expectation of $\sigma(t\bar{b}+\bar{t}b) = 11.07 ^{+0.19}_{-0.18}\,\si{\pico\barn}$. Beyond completing the experimental observation of all three single top-quark production mechanisms at the LHC, this result provides a crucial, independent validation of the electroweak sector. By establishing a robust measurement of this suppressed process, it opens a new phase space to directly probe the top quark's electroweak couplings and for constraining physics beyond the Standard Model. In particular, it offers sensitivity to $s$-channel heavy resonances, such as $W'$ or charged Higgs bosons~\cite{Tait:2000sh,Boos:2006xe}, and specific anomalous couplings via effective field theories~\cite{Zhang:2016omx}.


\FloatBarrier
\section*{Acknowledgements}
%

%

%
%

%
%

We thank CERN for the very successful operation of the LHC and its injectors, as well as the support staff at
CERN and at our institutions worldwide without whom ATLAS could not be operated efficiently.

The crucial computing support from all WLCG partners is acknowledged gratefully, in particular from CERN, the ATLAS Tier-1 facilities at TRIUMF/SFU (Canada), NDGF (Denmark, Norway, Sweden), CC-IN2P3 (France), KIT/GridKA (Germany), INFN-CNAF (Italy), NL-T1 (Netherlands), PIC (Spain), RAL (UK) and BNL (USA), the Tier-2 facilities worldwide and large non-WLCG resource providers. Major contributors of computing resources are listed in Ref.~\cite{ATL-SOFT-PUB-2026-001}.

We gratefully acknowledge the support of ANPCyT, Argentina; YerPhI, Armenia; ARC, Australia; BMWFW and FWF, Austria; ANAS, Azerbaijan; CNPq and FAPESP, Brazil; NSERC, NRC and CFI, Canada; CERN; ANID, Chile; CAS, MOST, NSFC and FRFCU, China; Minciencias, Colombia; MEYS CR, Czech Republic; DNRF and DNSRC, Denmark; IN2P3-CNRS and CEA-DRF/IRFU, France; SRNSFG, Georgia; BMFTR, HGF and MPG, Germany; GSRI, Greece; RGC and Hong Kong SAR, China; ICHEP and Academy of Sciences and Humanities, Israel; INFN, Italy; MEXT and JSPS, Japan; CNRST, Morocco; NWO, Netherlands; RCN, Norway; MNiSW, Poland; FCT, Portugal; MNE/IFA, Romania; MSTDI, Serbia; MSSR, Slovakia; ARIS and MVZI, Slovenia; DSI/NRF, South Africa; MICIU/AEI, Spain; SRC and Wallenberg Foundation, Sweden; SERI, SNSF and Cantons of Bern and Geneva, Switzerland; NSTC, Taipei; TENMAK, T\"urkiye; STFC/UKRI, United Kingdom; DOE and NSF, United States of America.

Individual groups and members have received support from BCKDF, CANARIE, CRC and DRAC, Canada; CERN-CZ and PRIMUS, Czech Republic; COST, ERC, ERDF, Horizon 2020 and Marie Sk{\l}odowska-Curie Actions, European Union; Investissements d'Avenir Labex, Investissements d'Avenir Idex and ANR, France; DFG and AvH Foundation, Germany; Herakleitos, Thales and Aristeia programmes co-financed by EU-ESF and the Greek NSRF, Greece; BSF-NSF and MINERVA, Israel; NCN and NAWA, Poland; La Caixa Banking Foundation, CERCA and AGAUR programs from Generalitat de Catalunya and PROMETEO and GenT Programmes Generalitat Valenciana, Spain; G\"{o}ran Gustafssons Stiftelse, Sweden; The Royal Society and Leverhulme Trust, United Kingdom; Eric and Wendy Schmidt Fund for Strategic Innovation, United States of America.

In addition, individual members wish to acknowledge support from Chile: Agencia Nacional de Investigaci\'on y Desarrollo (ANID FONDECYT reg. 1230987, FONDECYT 1230812, FONDECYT 1240864, Fondecyt 3240661, Fondecyt Regular 1240721); China: Fundamental Research Funds for the Central Universities (010-63263105), Chinese Ministry of Science and Technology (MOST-2023YFA1605700, MOST-2023YFA1609300), National Natural Science Foundation of China (NSFC 12275265, NSFC-W2543005); Czech Republic: Czech Science Foundation (GACR - 24-11373S), Ministry of Education Youth and Sports (ERC-CZ-LL2327, FORTE CZ.02.01.01/00/22\_008/0004632); EU: H2020 European Research Council (ERC - 101002463); European Union: European Research Council (BARD No. 101116429, ERC - 101219398, ERC 101089007), European Regional Development Fund (HE COFUND GA No.101081355, ERDF), Marie Sklodowska-Curie Actions (GAP-101168829); France: Agence Nationale de la Recherche (ANR-22-EDIR-0002, ANR-24-CE31-0504-01); China: Research Grants Council (GRF); Italy: Istituto Nazionale di Fisica Nucleare (LHC-MIUR - 28003/2025), Ministero dell'Università e della Ricerca (NextGenEU I53D23000820006 M4C2.1.1, SOE2024\_0000023); Japan: Japan Society for the Promotion of Science (JSPS KAKENHI  JP25H0063, JSPS KAKENHI JP22H04944, JSPS KAKENHI JP24K23939, JSPS KAKENHI JP24KK0251, JSPS KAKENHI JP25H00650, JSPS KAKENHI JP25H01291, JSPS KAKENHI JP25K01011, JSPS KAKENHI JP25K01023, JSPS KAKENHI JP25KK0047); Poland: Polish National Science Centre (NCN OPUS 2023/51/B/ST2/02507, NCN OPUS nr 2022/47/B/ST2/03059, UMO-2019/34/E/ST2/00393, UMO-2021/42/E/ST2/00350, UMO-2022/47/O/ST2/00148, UMO-2023/49/B/ST2/04085, UMO-2023/51/B/ST2/00920, UMO-2024/53/N/ST2/00869); Spain: Agència de Gestió d'Ajuts Universitaris i de Recerca. (AGAUR - 2023 BP 00141), Ministry of Science and Innovation (RYC2019-028510-I, RYC2020-030254-I, RYC2021-031273-I, RYC2022-038164-I), Ministerio de Ciencia, Innovación y Universidades/Agencia Estatal de Investigaci\'on (PID2022-142604OB-C22); Sweden: Carl Trygger Foundation (Carl Trygger Foundation CTS 22:2312), Swedish Research Council (Swedish Research Council 2023-04654, VR 2021-03651, VR 2022-03845, VR 2022-04683, VR 2023-03403, VR 2024-05451, VR 2025-05940), Knut and Alice Wallenberg Foundation (KAW 2023.0366); United Kingdom: The Binks Trust, Royal Society (NIF-R1-231091); United States of America: U.S. Department of Energy (ECA DE-AC02-76SF00515), John Templeton Foundation (John Templeton Foundation 63206), Neubauer Family Foundation.

%
%


%
%
%
%
%
%
%
%
%

%
\clearpage
\appendix
\printbibliography
\clearpage
 
\begin{flushleft}
\hypersetup{urlcolor=black}
{\Large The ATLAS Collaboration}

\bigskip

\AtlasOrcid[0000-0002-6665-4934]{G.~Aad}$^\textrm{\scriptsize 102}$,
\AtlasOrcid[0000-0001-7616-1554]{E.~Aakvaag}$^\textrm{\scriptsize 17}$,
\AtlasOrcid[0000-0002-5888-2734]{B.~Abbott}$^\textrm{\scriptsize 121}$,
\AtlasOrcid[0000-0002-0287-5869]{S.~Abdelhameed}$^\textrm{\scriptsize 83b}$,
\AtlasOrcid[0000-0002-1002-1652]{K.~Abeling}$^\textrm{\scriptsize 54}$,
\AtlasOrcid[0000-0001-5763-2760]{N.J.~Abicht}$^\textrm{\scriptsize 48}$,
\AtlasOrcid[0000-0002-8496-9294]{S.H.~Abidi}$^\textrm{\scriptsize 30}$,
\AtlasOrcid[0009-0003-6578-220X]{M.~Aboelela}$^\textrm{\scriptsize 44}$,
\AtlasOrcid[0000-0002-9987-2292]{A.~Aboulhorma}$^\textrm{\scriptsize 36e}$,
\AtlasOrcid[0000-0001-5329-6640]{H.~Abramowicz}$^\textrm{\scriptsize 154}$
% ...
\bigskip
\\

% ...
$^{17}$Department for Physics and Technology, University of Bergen, Bergen; Norway.\\
% ...
$^{30}$Physics Department, Brookhaven National Laboratory, Upton NY; United States of America.\\
% ...
$^{36}$$^{(a)}$Facult\'e des Sciences Ain Chock, Universit\'e Hassan II de Casablanca;$^{(b)}$Facult\'{e} des Sciences, Universit\'{e} Ibn-Tofail, K\'{e}nitra;$^{(c)}$Facult\'e des Sciences Semlalia, Universit\'e Cadi Ayyad, LPHEA-Marrakech;$^{(d)}$LPMR, Facult\'e des Sciences, Universit\'e Mohamed Premier, Oujda;$^{(e)}$Facult\'e des sciences, Universit\'e Mohammed V, Rabat;$^{(f)}$Institute of Applied Physics, Mohammed VI Polytechnic University, Ben Guerir; Morocco.\\
% ...
$^{44}$Physics Department, Southern Methodist University, Dallas TX; United States of America.\\
% ...
$^{48}$Fakult\"{a}t Physik, Technische Universit{\"a}t Dortmund, Dortmund; Germany.\\
% ...
$^{54}$II. Physikalisches Institut, Georg-August-Universit\"{a}t G\"ottingen, G\"ottingen; Germany.\\
% ...
$^{83}$$^{(a)}$Khalifa University of Science and Technology, Abu Dhabi;$^{(b)}$New York University Abu Dhabi, Abu Dhabi;$^{(c)}$United Arab Emirates University, Al Ain;$^{(d)}$University of Sharjah, Sharjah; United Arab Emirates.\\
% ...
$^{102}$CPPM, Aix-Marseille Universit\'e, CNRS/IN2P3, Marseille; France.\\
% ...
$^{121}$Homer L. Dodge Department of Physics and Astronomy, University of Oklahoma, Norman OK; United States of America.\\
% ...
$^{154}$Raymond and Beverly Sackler School of Physics and Astronomy, Tel Aviv University, Tel Aviv; Israel.\\
% ...
$^{a}$ Also at Affiliated with an institute formerly covered by a cooperation agreement with CERN.\\
$^{b}$ Also at An-Najah National University, Nablus; Palestine.\\
$^{c}$ Also at Borough of Manhattan Community College, City University of New York, New York NY; United States of America.\\
$^{d}$ Also at Center for Interdisciplinary Research and Innovation (CIRI-AUTH), Thessaloniki; Greece.\\
$^{e}$ Also at Centre of Physics of the Universities of Minho and Porto (CF-UM-UP); Portugal.\\
$^{f}$ Also at CERN, Geneva; Switzerland.\\
% ...

\end{flushleft}


%

%
%

%

%
%
%
%

\end{document}